
%
\magnification=1200
%
%
\catcode`\@=11 
\newcount\yearltd\yearltd=\year\advance\yearltd by -1900
%

\def\draftmode{\message{ DRAFTMODE }\def\draftdate{{\rm preliminary draft:
\number\month/\number\day/\number\yearltd\ \ \hourmin}}%
\headline={\hfil\draftdate}\writelabels\baselineskip=20pt plus 2pt minus 2pt
 {\count255=\time\divide\count255 by 60 \xdef\hourmin{\number\count255}
  \multiply\count255 by-60\advance\count255 by\time
  \xdef\hourmin{\hourmin:\ifnum\count255<10 0\fi\the\count255}}}
\def\nolabels{\def\wrlabeL##1{}\def\eqlabeL##1{}\def\reflabeL##1{}}
\def\writelabels{\def\wrlabeL##1{\leavevmode\vadjust{\rlap{\smash%
{\line{{\escapechar=` \hfill\rlap{\sevenrm\hskip.03in\string##1}}}}}}}%
\def\eqlabeL##1{{\escapechar-1\rlap{\sevenrm\hskip.05in\string##1}}}%
\def\reflabeL##1{\noexpand\llap{\noexpand\sevenrm\string\string\string##1}}}
\nolabels
%
\global\newcount\secno \global\secno=0
\global\newcount\meqno \global\meqno=1
\def\newsec#1{\global\advance\secno by1\message{(\the\secno. #1)}
\global\subsecno=0\eqnres@t\noindent{\bf\the\secno. #1}
\writetoca{{\secsym} {#1}}\par\nobreak\medskip\nobreak}
\def\eqnres@t{\xdef\secsym{\the\secno.}\global\meqno=1\bigbreak\bigskip}
\def\sequentialequations{\def\eqnres@t{\bigbreak}}\xdef\secsym{}
\global\newcount\subsecno \global\subsecno=0
\def\subsec#1{\global\advance\subsecno by1\message{(\secsym\the\subsecno. #1)}
\ifnum\lastpenalty>9000\else\bigbreak\fi
\noindent{\it\secsym\the\subsecno. #1}\writetoca{\string\quad
{\secsym\the\subsecno.} {#1}}\par\nobreak\medskip\nobreak}
\def\appendix#1#2{\global\meqno=1\global\subsecno=0\xdef\secsym{\hbox{#1.}}
\bigbreak\bigskip\noindent{\bf Appendix #1. #2}\message{(#1. #2)}
\writetoca{Appendix {#1.} {#2}}\par\nobreak\medskip\nobreak}
%
%
\def\eqnn#1{\xdef #1{(\secsym\the\meqno)}\writedef{#1\leftbracket#1}%
\global\advance\meqno by1\wrlabeL#1}
\def\eqna#1{\xdef #1##1{\hbox{$(\secsym\the\meqno##1)$}}
\writedef{#1\numbersign1\leftbracket#1{\numbersign1}}%
\global\advance\meqno by1\wrlabeL{#1$\{\}$}}
\def\eqn#1#2{\xdef #1{(\secsym\the\meqno)}\writedef{#1\leftbracket#1}%
\global\advance\meqno by1$$#2\eqno#1\eqlabeL#1$$}
%
%
\global\newcount\refno \global\refno=1
\newwrite\rfile
\def\ref{$^{\the\refno}$\nref}
\def\nref#1{\xdef#1{[\the\refno]}\writedef{#1\leftbracket#1}%
\ifnum\refno=1\immediate\openout\rfile=refs.tmp\fi
\global\advance\refno by1\chardef\wfile=\rfile\immediate
\write\rfile{\noexpand\item{#1\ }\reflabeL{#1\hskip.31in}\pctsign}\findarg}
\def\findarg#1#{\begingroup\obeylines\newlinechar=`\^^M\pass@rg}
{\obeylines\gdef\pass@rg#1{\writ@line\relax #1^^M\hbox{}^^M}%
\gdef\writ@line#1^^M{\expandafter\toks0\expandafter{\striprel@x #1}%
\edef\next{\the\toks0}\ifx\next\em@rk\let\next=\endgroup\else\ifx\next\empty%
\else\immediate\write\wfile{\the\toks0}\fi\let\next=\writ@line\fi\next\relax}}
\def\striprel@x#1{} \def\em@rk{\hbox{}}
\def\lref{\begingroup\obeylines\lr@f}
\def\lr@f#1#2{\gdef#1{\ref#1{#2}}\endgroup\unskip}
\def\semi{;\hfil\break}
\def\addref#1{\immediate\write\rfile{\noexpand\item{}#1}} 
\def\startrefs#1{\immediate\openout\rfile=refs.tmp\refno=#1}
\def\xref{\expandafter\xr@f}\def\xr@f[#1]{#1}
\def\cite{\expandafter\cxr@f}\def\cxr@f[#1]{$^{#1}$}
\def\refs#1{\count255=1$^{\r@fs #1{\hbox{}}}$}
\def\r@fs#1{\ifx\und@fined#1\message{reflabel \string#1 is undefined.}%
\nref#1{need to supply reference \string#1.}\fi%
\vphantom{\hphantom{#1}}\edef\next{#1}\ifx\next\em@rk\def\next{}%
\else\ifx\next#1\ifodd\count255\relax\xref#1\count255=0\fi%
\else#1\count255=1\fi\let\next=\r@fs\fi\next}
\newwrite\lfile
{\escapechar-1\xdef\pctsign{\string\%}\xdef\leftbracket{\string\{}
\xdef\rightbracket{\string\}}\xdef\numbersign{\string\#}}

\def\writestop{\def\writestoppt{\immediate\write\lfile{\string\pageno%
\the\pageno\string\startrefs\leftbracket\the\refno\rightbracket%
\string\def\string\secsym\leftbracket\secsym\rightbracket%
\string\secno\the\secno\string\meqno\the\meqno}\immediate\closeout\lfile}}
\def\writestoppt{}\def\writedef#1{}
\def\seclab#1{\xdef #1{\the\secno}\writedef{#1\leftbracket#1}\wrlabeL{#1=#1}}
\def\subseclab#1{\xdef #1{\secsym\the\subsecno}%
\writedef{#1\leftbracket#1}\wrlabeL{#1=#1}}
\newwrite\tfile \def\writetoca#1{}
\def\leaderfill{\leaders\hbox to 1em{\hss.\hss}\hfill}
\def\writetoc{\immediate\openout\tfile=toc.tmp
   \def\writetoca##1{{\edef\next{\write\tfile{\noindent ##1
   \string\leaderfill {\noexpand\number\pageno} \par}}\next}}}

%
\catcode`\@=12 
%
%
\font\abssl=cmsl10 scaled 833
\font\absrm=cmr10 scaled 833 \font\absrms=cmr7 scaled  833
\font\absrmss=cmr5 scaled  833 \font\absi=cmmi10 scaled  833
\font\absis=cmmi7 scaled  833 \font\absiss=cmmi5 scaled  833
\font\abssy=cmsy10 scaled  833 \font\abssys=cmsy7 scaled  833
\font\abssyss=cmsy5 scaled  833 \font\absbf=cmbx10 scaled 833
\skewchar\absi='177 \skewchar\absis='177 \skewchar\absiss='177
\skewchar\abssy='60 \skewchar\abssys='60 \skewchar\abssyss='60
\def\abstractfont{\def\rm{\fam0\absrm}
\textfont0=\absrm \scriptfont0=\absrms \scriptscriptfont0=\absrmss
\textfont1=\absi \scriptfont1=\absis \scriptscriptfont1=\absiss
\textfont2=\abssy \scriptfont2=\abssys \scriptscriptfont2=\abssyss
\textfont\itfam=\bigit \def\it{\fam\itfam\bigit}\def\footnotefont{\tenpoint}%
\textfont\slfam=\abssl \def\sl{\fam\slfam\abssl}%
\textfont\bffam=\absbf \def\bf{\fam\bffam\absbf}\rm}
\font\ftsl=cmsl10 scaled 833
\font\ftrm=cmr10 scaled 833 \font\ftrms=cmr7 scaled  833
\font\ftrmss=cmr5 scaled  833 \font\fti=cmmi10 scaled  833
\font\ftis=cmmi7 scaled  833 \font\ftiss=cmmi5 scaled  833
\font\ftsy=cmsy10 scaled  833 \font\ftsys=cmsy7 scaled  833
\font\ftsyss=cmsy5 scaled  833 \font\ftbf=cmbx10 scaled 833
\skewchar\fti='177 \skewchar\ftis='177 \skewchar\ftiss='177
\skewchar\ftsy='60 \skewchar\ftsys='60 \skewchar\ftsyss='60
\def\footnotefont{\def\rm{\fam0\ftrm}
\textfont0=\ftrm \scriptfont0=\ftrms \scriptscriptfont0=\ftrmss
\textfont1=\fti \scriptfont1=\ftis \scriptscriptfont1=\ftiss
\textfont2=\ftsy \scriptfont2=\ftsys \scriptscriptfont2=\ftsyss
\textfont\itfam=\bigit \def\it{\fam\itfam\bigit}\def\footnotefont{\tenpoint}%
\textfont\slfam=\ftsl \def\sl{\fam\slfam\ftsl}%
\textfont\bffam=\ftbf \def\bf{\fam\bffam\ftbf}\rm}
%
\def\tenpoint{\def\rm{\fam0\tenrm}
\textfont0=\tenrm \scriptfont0=\sevenrm \scriptscriptfont0=\fiverm
\textfont1=\teni  \scriptfont1=\seveni  \scriptscriptfont1=\fivei
\textfont2=\tensy \scriptfont2=\sevensy \scriptscriptfont2=\fivesy
\textfont\itfam=\tenit \def\it{\fam\itfam\tenit}\def\footnotefont{\ninepoint}%
\textfont\bffam=\tenbf \def\bf{\fam\bffam\tenbf}\def\sl{\fam\slfam\tensl}\rm}
\font\ninerm=cmr9 \font\sixrm=cmr6 \font\ninei=cmmi9 \font\sixi=cmmi6
\font\ninesy=cmsy9 \font\sixsy=cmsy6 \font\ninebf=cmbx9
\font\nineit=cmti9 \font\ninesl=cmsl9 \skewchar\ninei='177
\skewchar\sixi='177 \skewchar\ninesy='60 \skewchar\sixsy='60
\def\ninepoint{\def\rm{\fam0\ninerm}
\textfont0=\ninerm \scriptfont0=\sixrm \scriptscriptfont0=\fiverm
\textfont1=\ninei \scriptfont1=\sixi \scriptscriptfont1=\fivei
\textfont2=\ninesy \scriptfont2=\sixsy \scriptscriptfont2=\fivesy
\textfont\itfam=\ninei \def\it{\fam\itfam\nineit}\def\sl{\fam\slfam\ninesl}%
\textfont\bffam=\ninebf \def\bf{\fam\bffam\ninebf}\rm}
%
%
\def\tr{{\rm tr}} 
\font\bigit=cmti10 scaled \magstep1
\vsize=8.5truein
\hsize=6truein
\baselineskip 14truept plus 0.5truept minus 0.5truept
\def\ie{{\it i.e.}}
\def\tr{\,{\hbox{tr}}\,}
\def\eg{{\it e.g.}}
\def\neath#1#2{\mathrel{\mathop{#1}\limits_{#2}}}
\def\lsim{\mathrel{\rlap{\lower4pt\hbox{\hskip1pt$\sim$}}
    \raise1pt\hbox{$<$}}}         
\def\gsim{\mathrel{\rlap{\lower4pt\hbox{\hskip1pt$\sim$}}
    \raise1pt\hbox{$>$}}}         
%
%
\tolerance=10000
\hfuzz=5pt
\pageno=0\nopagenumbers\tolerance=10000\hfuzz=5pt
\line{\hfill CERN-TH.7453/94}
\vskip 24pt
\centerline{\bf POLARIZED STRUCTURE FUNCTIONS:}
\centerline{\bf THEORY AND PHENOMENOLOGY}
\vskip 36pt\centerline{Stefano~Forte\footnote*{On leave
from INFN, Sezione di Torino, Turin, Italy.}}
\vskip 12pt
\centerline{\it Theory Division, CERN,}
\centerline{\it CH-1211 Gen\`eve 23, Switzerland.}
\vskip 48pt
{\narrower\baselineskip 10pt
\centerline{\bf Abstract}
\medskip
We review the theory of polarized structure functions
measured in deep-inelastic lepton-nucleon scattering, focusing on the
most recent developments. We
concentrate on the structure function $
g_1$,
emphasizing the phenomenological problems related to the extraction of
the proton and neutron $g_1(x,Q^2)$ from the data,
and to the determination of its moments, especially the first moment.
In particular,  we discuss the theoretical and experimental
uncertainties due to the  $Q^2$ dependence of the
data, small and large $x$ extrapolations, QCD loop corrections,
higher twist corrections, and, in connection to neutron experiments,
nuclear effects. We also discuss the current status of the sum rules
satisfied by the first moment of $g_1$ and their theoretical
interpretation.\smallskip}
\vskip 50pt
\centerline{Invited talk at the}
\centerline{\it Tennessee International Symposium on Radiative
Corrections}
\centerline{Gatlinburg, Tennessee, June 1994}
\medskip
\centerline{\it to be published in the proceedings}
\vskip 55pt
\line{CERN-TH.7453/94\hfill}
\line{September 1994\hfill}
\eject
\footline={\hss\tenrm\folio\hss}
\centerline{ POLARIZED STRUCTURE FUNCTIONS:}
\centerline{ THEORY AND PHENOMENOLOGY}
\bigskip
\centerline{Stefano Forte\footnote*{\footnotefont On
leave from INFN, Sezione di
Torino, Italy}}
\smallskip
\centerline{\it Theory Division, CERN}
\centerline{\it CH-1211 Gen\`eve 23, Switzerland}
\bigskip
\centerline{ ABSTRACT}
\medskip
{\abstractfont\baselineskip 12 truept
\advance\leftskip by 36truept\advance\rightskip by 36truept
We review the theory of polarized structure functions
measured in deep-inelastic lepton-nucleon scattering, focusing on the
most recent developments. We
concentrate on the structure function $
g_1$,
emphasizing the phenomenological problems related to the extraction of
the proton and neutron $g_1(x,Q^2)$ from the data,
and to the determination of its moments, especially the first moment.
In particular,  we discuss the theoretical and experimental
uncertainties due to the  $Q^2$ dependence of the
data, small and large $x$ extrapolations, QCD loop corrections,
higher twist corrections, and, in connection to neutron experiments,
nuclear effects. We also discuss the current status of the sum rules
satisfied by the first moment of $g_1$ and their theoretical
interpretation.
\smallskip}

\baselineskip 14truept plus 0.5truept minus 0.5truept
\medskip
\goodbreak
\newsec{\bf Introduction}
\nobreak
Polarized deep-inelastic scattering provides a handle on the matrix
elements of spin-dependent operators. Because of the intricacies of
spin physics, this poses  exacting challenges to
experimentalists and theorists alike. On the theoretical side, it turns out
to be equally hard to extract meaningful physical information from the
data, and to understand the significance of that information.

This is perhaps best appreciated by recalling the present status of
the measurement of the first moment of the polarized structure
function $g_1$:
\eqn\gamdef
{\Gamma_1=\int_0^1\! dx\, g_1(x)
.}
Among polarized observables, this is the simplest to understand
theoretically because there exists only one operator with the
appropriate spin at leading twist in the operator-product expansion,
the flavor singlet fermionic axial current
\eqn\acurr
{j^\mu_5=\sum_{i=1}^{N_f} \bar \psi_i \gamma_\mu \gamma_5 \psi_i.}
A measurement of $\Gamma_1$  thus determines
(in a way to be discussed extensively below) the (forward) matrix element of
$j^\mu_5$ in the target:
\eqn\jmel
{\langle p, s |j^\mu_5| p, s \rangle=  M s^\mu \Delta \Sigma,}
where $p^\mu$, $M$, and $s^\mu$ are, respectively,  the target
four-momentum, mass, and spin (normalized as $ s^\mu s_\mu=-1$).

The experimental  value of $\Delta \Sigma$ for a proton target,
as given by the various
experimental collaborations which have measured it, is displayed in
Fig.~1a.\nref\EMCa{EMC Collaboration,
J.~Ashman et al., {\it Phys. Lett.} {\bf B206} (1988)
364}\nref\EMCb{EMC
Collaboration,
J.~Ashman et al., {\it Nucl. Phys.} {\bf B328} (1989) 1}\nref\wa{SMC
Collaboration,
B.~Adeva et al., {\it Phys. Lett.} {\bf B320} (1994) 400}\nref\SMC{SMC
Collaboration,
D.~Adams et al., {\it Phys. Lett.} {\bf B329} (1994)
399}\nref\SLAC{E143
Collaboration, R.~Arnold et al.,
preliminary results presented at ICHEP94, Glasgow,
August 1994}\nref\stuart{L.~Stuart, these
proceedings}\nref\JM{R.~L.~Jaffe
and A.~Manohar, {\it Nucl. Phys.} {\bf B337} (1990)
509}\nref\EK{J.~Ellis
and M.~Karliner,
{\it Phys. Lett.} {\bf B313} (1993) 131;
talk at PANIC 1993, Perugia, July 1993, preprint
CERN-TH.7022/93 }\nref\CR{F.~Close and R.~D.~Roberts, {\it Phys. Lett.}
{\bf B316} (1993)165}  All the
determinations are compatible within errors; however,
the spread of the results
indicates how hard these measurements are, and how
carefully they should be taken. Actually, the experimental result
which has been around longest\cite\EMCb\
has been reanalysed by various
theoretical groups: the results, shown in Fig.~1b, again display a
remarkable degree of variation (also in the error estimate).
Finally, it is  clear (albeit with hindsight) that
whatever the measured value of $\Delta \Sigma$, its interpretation will not be
obvious: the current $j_5^\mu$ is not conserved because of the
axial anomaly, hence  its matrix elements do not directly measure
a well-defined conserved quantum number of a physical state.

The increase in experimental precision calls now for a careful
reanalysis
of the  arguments which go into the
extraction
of structure functions from the data: indeed, the theoretical
uncertainty  starts to be comparable to the
experimental error. On the other hand,
after the considerable amount of
theoretical work which followed the 1988 EMC experiment,\cite\EMCb\ the
 subtleties involved in the interpretation of the
results displayed in Fig.~1 are now largely understood.

In this paper we will concentrate on the phenomenological aspects of
structure function measurements, with particular regard to recent
developments,  and we will briefly summarize the present understanding
of their theoretical import. In Sect.~2 we will provide some
 background on the polarized structure functions
$g_1(x,Q^2)$ and $g_2(x,Q^2)$.
In Sect.~3 we will then proceed to a discussion of the phenomenology of
the proton structure function $g_1^p$: first we will discuss the way
$g_1$ is extracted from the data, and review the problems involved in
$Q^2$ dependence and extrapolation in $x$; then we will examine
the way $\Delta \Sigma$ [Eq.~\jmel] is evaluated once $g_1$ is known,
discussing the dependence on weak decay constants and the role of QCD
loop corrections. In Sect.~4 we will  review the current status of the
theoretical interpretation of the data on $\Delta \Sigma$, and we will
discuss parametrization of polarized parton distributions. In Sect.~5
we will review problems which are of special relevance for
the measurement of the
neutron structure function $g_1^n$, in particular nuclear effects
and higher twist effects, and in Sect.~6 we
will discuss the significance of the proton and neutron
measurements taken together.\footnote*{\footnotefont\baselineskip 11 truept
For a more detailed introduction to
the theory of $g_1$  in QCD and to theoretical controversies spurred by
its measurement the reader is referred to earlier
reviews,\nref\guidoer{G.~Altarelli, in ``The challenging questions'',
Proc. of the 1989 Erice School,
A.~Zichichi, ed. (Plenum, New York, 1990)}\refs{\JM,\guidoer} while
a recent
assessment of the theoretical status of the $g_1$ measurement is in
\nref\guidomo{G.~Altarelli and G.~Ridolfi,
preprint CERN-TH.7415/94}Ref.~\xref\guidomo; for a detailed review of the
theory of $g_2$ see Ref.~\nref\jagt{R.~L.~Jaffe, {\it Comm. Nucl. Part.
Phys.}
{\bf 14} (1990) 239}\xref\jagt, and for a discussion
of other polarized structure functions (relevant for different processes,
such as Drell-Yan scattering) Ref.~\nref\jah{R.~L.~Jaffe, in
``Structure of baryons and related mesons'', Proc. of the Baryon'92
Conference, M.~Gai, ed. (World Scientific, Singapore, 1993)}\xref\jah.}
\goodbreak
\medskip
\newsec{\bf The structure functions and their moments}
\nobreak
Polarized  structure functions are the form factors which parametrize
the cross section spin asymmetry
for deep-inelastic scattering of polarized leptons
off a polarized hadronic target
${d^2(\sigma^{\uparrow\uparrow}-\sigma^{\uparrow\downarrow})\over d Q^2 d\nu}$;
they are given by the antisymmetric part of the hadronic
tensor\nref\guidopr{G.~Altarelli, {\it Phys. Rep.}
{\bf 81} (1982)  1}\nref\koda{J.~Kodaira et al., {\it Phys.
Rev.} {\bf D20} (1979) 627; {\it Nucl. Phys.}
{\bf B159} (1979) 99}\refs{\koda,\guidopr}
\eqn\wdef
{\eqalign{i W_A^{\mu\nu}&\equiv {1\over 4\pi}\int\! d^4x\,
e^{i q\cdot x} \langle p, s |
J^{[\mu}(x) J^{\nu]}(0) | p,s\rangle
\cr
&=i M \epsilon^{\nu\nu\rho\sigma} q_\rho \left[ {s_\sigma\over p\cdot q}
g_1(x, Q^2)+ {s_\sigma p\cdot q - p_\sigma q\cdot s\over(p\cdot q)^2}
g_2(x, Q^2)\right]\cr}}
(with standard parton model
kinematics),
where $M$ is the target mass. Polarized neutrino scattering is only of
academic interest at present, hence  we will neglect weak interaction
effects, and assume
$J^\mu$ in Eq.~\wdef\ to be electric currents.

The light-cone expansion of the current product in Eq.~\wdef\
implies
that the moments of the structure
functions
$g_1$ and $g_2$ in the Bjorken limit are given in the nonsinglet case
by\refs{\koda,\guidopr}
\eqnn\opconta\eqnn\opcontb
$$
\eqalignno{\int_0^1\! dx \, x^{n-1} g_1(x, Q^2) &= {1\over 2}C^n_1 (Q^2) a^n
(Q^2)&\opconta\cr
\int_0^1\! dx \, x^{n-1} g_2(x, Q^2) &= {n-1\over2 n}\left[C^n_2 (Q^2) d^n
(Q^2)-C^n_1 (Q^2) a^n (Q^2)\right].&\opcontb\cr}$$
$C_n(Q^2)$ are perturbatively calculable coefficient functions,
and $a_n$ and $d_n$ are given by  matrix elements of the leading twist
polarized operators:
\eqnn\ltopa\eqnn\ltopb
$$\eqalignno{
M a_n s^{\{\sigma} p^{\mu_1}\dots p^{\mu_{n-1}\}}
&=-\langle p, s | i^{n-1} \bar
\psi\gamma_5\gamma^{\{\sigma}D^\mu_1\dots D^{\mu_{n-1}\}}\lambda_i\psi
|ps\rangle & \ltopa\cr
M d_n s^{\{\sigma} p^{\mu_1}\dots p^{\mu_{n-1}\}}
&=\langle p, s | i^{n-1} \bar
\psi\gamma_5\gamma^{\{[\sigma}D^{\mu_1]}\dots
D^{\mu_{n-1}\}}\lambda_i\psi|ps\rangle, & \ltopb\cr}
$$
where $\{\}$  denotes complete symmetrization. Notice that, even though
the operators Eq.~\opconta\ are twist-2 and the operators Eq.~\opcontb\
are twist-3, their respective contributions to
the light-cone expansion of $W^{\mu\nu}$ are  of the same order in
$Q^2$. In the singlet case, these operators will further mix with
gluonic ones;\footnote*{\footnotefont\baselineskip 11 truept
Note also that Eq.~\ltopb\ is true only for strictly
massless quarks; in general the operators contributing to $g_2$ will
mix with twist-3 mass dependent operators. The contribution of these
operators does not vanish asymptotically, even though their matrix
elements should be of order $m/M$, with $m$ and $M$ the quark and
nucleon masses. }\ref\kodsolo{J.~Kodaira, {\it Nucl. Phys.}
{\bf B165} (1980)129} however, the first moment of $g_1$, which we will be
mostly concerned with, is still given by Eq.~\ltopa.

The reason why we will concentrate our discussion on
$g_1$ is that for longitudinally
polarized protons the only nonvanishing components of the
hadronic tensor  are\cite\guidopr\ (in the frame where, say, only the third
spatial component of $q$ is nonzero and its energy vanishes)
\eqn\lonpol
{W^{12}_A=-W^{21}_A=\pm
\left[ g_1 -\left({2 Mx\over Q}\right)^2g_2\right],}
hence only $g_1$ is
relevant asymptotically. For transverse polarization (along, say, the
first spatial coordinate)
the nonvanishing components are instead
\eqn\tranpol
{W^{02}_A=-W^{20}_A={2Mx\over Q}\left[g_1+g_2\right],}
thus, even
though the two structure functions contribute equally to it, the whole
cross section vanishes asymptotically. Only an experimental
upper bound on $g_2$ is thus presently available;\ref\SMCtran{SMC
Collaboration,
D.~Adams et al., {\it Phys. Lett.} {\bf B336}, 125 (1994)} an
alternative
determination of it is
forthcoming.\cite\stuart

In general, $\Gamma_1$ will receive both singlet and nonsinglet
contributions, and it is therefore convenient to define axial charges
$a_i$ for the $i$-th flavor:
\eqn\fljmel
{M a_is^\mu\equiv \langle p, s |\bar \psi_i \gamma_\mu \gamma_5 \psi_i
| p, s \rangle.}
The first moment of $g_1$ [Eq.~\gamdef] for a nucleon  target  is then given by
\eqn\firstmom{
\eqalign{\Gamma^{p,\,n}_1(Q^2&)=
{1\over 2} \left[\sum_{i=1}^{N_f} e^2_i C_i(Q^2)
a_i\right]\cr
={1\over 12}&\left[C_{NS}(Q^2)\left[\pm\left(a_u-a_d\right)+{1\over
3}\left( a_u+a_d-2a_s\right)\right]+C_S(Q^2) {4\over
3}\left(a_u+a_d+a_s\right)
\right],\cr
}}
where in the last step we have assumed that
only the three lightest flavors are activated, the plus (minus) sign
refers to the proton (neutron), and the explicit form
of the singlet and nonsinglet coefficient functions $C_S$, $C_{NS}$
will be given in Sect.~3.2.
\goodbreak
\medskip
\newsec{\bf Phenomenology of $g_1^p$}
\nobreak
The primary quantity from which experimental information on polarized
structure functions is obtained is the polarization
asymmetry
$A(x,Q^2)={ \sigma^{\uparrow \uparrow}- \sigma^{\uparrow
\downarrow}\over  \sigma^{\uparrow \uparrow}+ \sigma^{\uparrow \downarrow}}$.
In order to relate this to $W^{\mu\nu}$ [Eq.~\wdef],
we must extract from $A$  the
asymmetry
for scattering of transverse (\ie\ with longitudinal helicity)
virtual photons, defined as
\eqn\asonedef{
A_1(x,Q^2)= { \sigma_{1/2}- \sigma_{3/2}\over
 \sigma_{1/2}+ \sigma_{3/2}}}
where the subscripts denote the total angular momentum of the
photon-nucleon pair along the incoming electron's direction; the
denominator of \asonedef\ is the total transverse
photoabsorption cross section $\sigma_T=\sigma_{1/2}+\sigma_{3/2}$.
$A$ is related to $A_1$ by
$A= D(A_1+\eta A_2)$,
in terms of the factor $D$ which provides the longitudinal
depolarization of the virtual photon with respect to its parent lepton,
the interference $A_2$ with the longitudinal photon polarization
amplitude, and a kinematic factor $\eta$.

The asymmetry $A_1$ measures  the ratio
between the  combination \wdef\ of  polarized structure functions, and
the unpolarized structure function $F_1$ (the other independent unpolarized
structure function gives the total longitudinal cross section).
In the Bjorken limit the contribution of
$g_2$ is negligible [Eq.~\lonpol] only $g_1$
contributes to the numerator in Eq.~\asonedef, and $g_1$ is given by
\eqnn\afone\eqnn\aftwo
$$\eqalignno{
g_1(x,Q^2)&= A_1(x,Q^2) F_1(x, Q^2)&\afone\cr
&= A_1(x,Q^2) {F_2(x, Q^2)\over 2x\left[1+R(x,Q^2)\right]},&\aftwo\cr}
$$
where the last step holds at leading twist.
The unpolarized
structure function $F_2$ is  directly determined from experiment,
and has the simple parton model interpretation (which in the DIS
scheme remains true to all orders in perturbative QCD)
\eqn\ftwo
{{F_2(x)\over x}=\sum_{i=1}^{N_f} e^2_i \left[q_i(x)+\bar
q_i(x)\right],}
while  $R={ \sigma_L\over  \sigma_T}$, where $\sigma_L$ is the
longitudinal virtual photoabsorption cross section.
Because in the Bjorken limit $R$ vanishes,  Eq.~\ftwo\ reduces
to
$g_1\approx A_1 F_2/(2x)$; in particular, in the naive parton model it
thus follows that
\eqn\naive
{g_1^{\rm parton}(x)=
\sum_{i=1}^{N_f} e^2_i \left[\Delta q_i(x)+
\Delta \bar q_i(x)\right],}
where $\Delta q_i$ are the polarized parton distributions for quarks
of flavor $i$.

A precise determination of $g_1$ thus hinges on the following assumptions.
First, $A_1$ must be determined from the measured asymmetry $A$: in
older experiments\refs{\EMCa,\EMCb} this was done by simply neglecting
$A_2$ and identifying $A$ with $DA_1$.
Then, the
contribution of $g_2$ to $A_1$ is neglected, exploiting Eq.~\lonpol, so
that Eq.~\afone\ holds. Finally, $g_1$  is determined from the
measured $F_2$ using Eq.~\aftwo, the main uncertainty being in the
experimental knowledge of $R$.
The first assumption is under control: $A_2$
satisfies a
positivity limit $|A_2|\leq \sqrt{R}$, and is now\cite\SMCtran\
measured to be
compatible with zero within uncertainties, so that (taking also
advantage of the fact that  the coefficient $\eta$ turns out to be
small in the kinematic range covered by present experiments) its
possible effects are just included in the systematic error.
The neglect of $g_2$ should not affect the determination of the first
moment of $g_1$, since, regardless of the size of $g_2$, its first
moment is expected to vanish (according to the Burkhardt-Cottingham
sum rule);\nref\bksr{G.~Altarelli, B.~Lampe, P.~Nason and G.~Ridolfi,
preprint CERN-TH.7254/94 (1994)}\refs{\jagt,\bksr} even if it did not (due to
nonperturbative effects) its contribution is
estimated\ref\mauro{M.~Anselmino and
E.~Leader, {\it Z. Phys.} {\bf C41} (1988) 239}
to modify the
first moment of $g_1$ by at most 1\%. A precise
measurement of $g_2$ would settle the issue. The last step
of the analysis is also under complete control: due to
cancellations in the
dependence of $D$ and $F_1$ on $R$, the
asymmetry $g_1$ is in fact essentially independent of $R$.\cite\SMC\
All these effects combined lead thus to an uncertainty of a few per cent,
which is included in the experimental error and is smaller than the
uncertainty on the unpolarized structure function $F_2$.

In the remainder of this section, we will consider in particular the
proton experiments, which afford the best experimental accuracy.
\smallskip
\subsec{\it Scale dependence and x-extrapolation}
The values of $g_1$ which can be determined using Eq.~\aftwo\ span in
practice  a limited kinematic range:
very small
and large $x$ regions are  excluded, while  low $x$
data are taken at low $Q^2$ and conversely. Thus, in order to
obtain a determination of $g_1(x,Q^2)$ for all $x$ and one single
value of $Q^2$, the data must be evolved to a common $Q^2$ and then
extrapolated over the whole $x$ range.

The scale dependence of the asymmetry $A_1$ is given by the
Altarelli-Parisi equations, and follows from the fact that $g_1$ and
$F_2$ depend on $Q^2$ in different ways, even  at leading order.
This is apparent if one considers the respective first
moments: at one loop the anomalous dimension of
$\Gamma_1$ vanishes (the axial current is classically conserved; this
conservation is spoilt by the axial anomaly, but this is a two-loop
effect), while the first moment of $F_2$ evolves since obviously the
total number of $q$-$\bar q$ pairs is not conserved.
The scale dependence is thus calculable, the only
theoretical uncertainty being that related to the relatively poor
knowledge of polarized parton distributions (see Sect.~4.2).
The result of the calculation,\ref\anr{G.~Altarelli, P.~Nason and
G.~Ridolfi, {\it Phys. Lett.} {\bf B320} (1994) 152} with two extreme
scenarios for the polarized gluon distribution, is displayed
and compared with the data of Ref.~\xref\EMCa\ in
Fig.~2a.\nref\SLACold{E80
Collaboration, M.~J.~Alguard et al., {\it Phys.
Rev. Lett.} {\bf 37} (1976) 1261; {\bf 41} (1978) 70\semi\
E130 Collaboration, G.~Baum et al., {\it Phys. Rev. Lett.} {\bf 51}
(1983) 1135}

Clearly, the data are not precise enough to display the calculated
scale dependence, even though they are consistent with it
(the same applies to more recent data\cite\wa), which
explains why the scale dependence of the asymmetry is not yet
observed directly. Nevertheless, the average scale of the smallest and
largest $x$ bins is often rather different: for example in Ref.~\xref\SMC\
the smallest $x$ bin
($\langle x\rangle =0.005$) has $\langle Q^2\rangle = 1.3$ GeV$^2$, and the
largest $x$ bin ($\langle x\rangle =0.48$) has
$\langle Q^2\rangle = 58.0$ GeV$^2$. It is therefore
important to correct the data for this effect, \ie\ to evolve the
measured asymmetry to the common scale at which $g_1$ is to be
determined. The size of the correction (shown in Fig.~2b for the data
of Ref.~\xref\EMCb) depends strongly on this
scale: the correction on $\Gamma^p_1$ is smaller than 5\% for
the determination of Ref.~\xref\EMCb, which is given at
$Q^2=10.7$~GeV$^2$, but is for instance
of order 10\% for the deuteron data\ref\SMCn{SMC Collaboration,
B.~Adeva et al., {\it Phys. Lett.} {\bf B302} (1993) 533} at
$Q^2=4.7$ GeV$^2$. The overall effect is anyhow
larger than the uncertainty involved in its
determination (which is mostly due to the uncertainty in the form of the
polarized gluon distribution) and should therefore be included in the
analysis of the data, especially if  $g_1$ is to be determined at
low ($Q^2\lsim 5$~GeV$^2$) scales.

Once the data are evolved to a common scale,
$g_1$ is still only known over a limited $x$ range: for example, the
combined data of
Refs.\xref\SLACold\ and \xref\EMCb\ have $0.01\le x\le 0.7$, and the more
recent ones of Ref.~\xref\SMC\ have $0.003\le x\le0.7$.
In order to be able to compute moments of $g_1$, which are the primary
quantities in QCD, $g_1$ thus has to be extrapolated at small and
large $x$.

The large $x$ behavior should be controlled by QCD counting
rules\nref\brodcou{R.~Blankenbecler and S.~J.~Brodsky, {\it Phys. Rev.}
{\bf D10} (1974) 2973; J.~F.~Gunion, {\it Phys. Rev.} {\bf D10} (1974)
242 }\nref\brobu{S.~J.~Brodsky, M.~Burkardt and I.~Schmidt, preprint
SLAC-PUB-6087 (1994)}:\refs{\brodcou,\brobu}
because for large $x$ the virtuality of
the struck quark is large, the scattering process should be purely
perturbative, and the dominant contribution to it should come
from minimally connected tree graphs, \ie\ those where three valence
quarks exchange two perturbative gluons. This leads to the prediction
\eqn\courule
{G_q\sim (1-x)^p;\quad p=2n-1+2\Delta s_z,}
where $G_q$ is any parton distribution, $n$ is the number of spectator
quark lines (\ie, $n=2$ for a nucleon) and $\Delta s_z=0$ ($\Delta
s_z=1$)
for  parallel (antiparallel) quark and proton helicities. This implies
that in the approximation of Eq.~\naive\ one would expect
\eqn\largex
{A_1\neath\sim{x\to1}{\rm const.}; \quad
g_1\neath\sim{x\to1} (x-1)^3.}

The fall-off of $g_1$ at large $x$ implies that the contribution of
the large $x$ extrapolation to $\Gamma_1$
is expected to be small, \ie\ around a few per cent of $\Gamma_1$. The
extrapolation has been performed by fitting a phenomenological function
either to  $A_1$ (which amounts to fitting a
straight line through the last few data points),\refs{\EMCb\SMC}  or to
$g_1$.\cite\SLAC\ The  (preliminary)
result of the latter extrapolation
actually turns out to be
surpisingly large (by a factor 3 larger than that of Ref.~\xref\SMC);
this
appears to be due to an actual difference in the data, and not to
the method used in the extrapolation; however,
the large $x$ points have large uncertainties
 ---   indeed all
experiments are compatible with each other ---
and the effect on $\Gamma_1$
is anyway modest. The same applies to results obtained\cite\brobu\ by
imposing the behavior Eq.~\courule. A different estimate has also been
obtained by using a valence quark model,\cite\CR which is consistent
with the behavior \courule\ but predicts also its normalization; the
results are similar (see however the discussion of the deuterium data
in Sect.~5).

The small-$x$ extrapolation is somewhat harder to control. Arguments
based on the dominance of known Regge
poles\ref\heim{R.~L.~Heimann, {\it Nucl. Phys.} {\bf B64} (1973)
429} lead to the expectation
\eqn\smallxre
{ g_1\neath \sim {x\to 0} x^\alpha;\quad 0\le\alpha\le 0.5.}
However, use of  perturbative QCD together with  an Ansatz for the
nucleon wave function\cite\brobu\ suggests that the gluon
distribution should behave as
\eqn\smallxbro
{{\Delta G_q\over G}\neath \sim {x\to 0} x;}
thus, if the unpolarized gluon distribution  is dominated by a soft
pomeron,\ref\smallx{See \eg\ B.~Bade\l ek et al., {\it Rev. Mod. Phys.} {\bf
64}, 927 (1992)}
then  $G(x)\sim{1\over x}$ so that  the lower bound for $\alpha$ in
Eq.~\smallxre\ is saturated; but if $G(x)$ has a harder behavior (such as
a supercritical or
perturbative pomeron\cite\smallx) then $\alpha<0$ ($\alpha\sim
-0.1$ with   a supercritical pomeron, and even lower with the Lipatov
hard pomeron\cite\smallx). Due to QCD evolution the gluon feeds into the
quark at small $x$ (the  eigenvector of the evolution being a linear
combination of $\Delta q $ and $\Delta g$), and a singular behavior
with the same value of $\alpha$ is then induced in $g_1$.\ref\bfr{R.~D.~Ball,
S.~Forte and G.~Ridolfi, ``Polarized structure functions at small $x$'',
{\it in preparation}}
Also, a model of the pomeron based on nonperturbative gluon
exchange\ref\bala{S.~D.~Bass and P.~V.~Landshoff, Cambridge preprint
DAMTP 94/50} leads to
the still singular but softer behavior
\eqn\smallxbala
{g_1\neath \sim {x\to 0}-2 \ln x.}
Finally, it has been argued\ref\CRb{F.~E.~Close and R.~G.~Roberts,
{\it Phys. Lett.} {\bf B336} (1994) 257}
that negative signature cuts could induce
an even more singular behavior
\eqn\smallxcloro
{g_1\neath \sim {x\to 0} {1\over x \ln^2 x},}
even though there is no compelling theoretical evidence in favor of such
contributions. It is  worth noticing that not all these
behaviors are stable upon perturbative evolution:\cite\bfr\ in general if,
at some scale $Q_0$, $g_1$
is regular, or has a singularity of the form \smallxre\ with
$\alpha\lsim 0.1$ or softer [such as \smallxbala],
then at larger $Q$ and small $x$, due to QCD evolution,
it will develop a double logarithmic singularity\ref\einh{M.~A.~Ahmed
and G.~G.~Ross, {\it Phys. Lett.} {\bf B56} (1975) 385\semi M.~B.~Einhorn
and J.~Soffer, {\it Nucl. Phys.} {\bf B74} (1986) 714}
\eqn\smallxdas
{g_1\neath \sim {x\to 0} \exp2\gamma\sqrt{\log 1/x\, \log\log
Q^2};\quad
\gamma^2=(16/33 - 2 n_f)[5+4\sqrt{1-3n_f/ 32}].}
More
singular behaviors such as \smallxcloro \ will instead be preserved by the
evolution. The issue is further complicated by the mixing of
quark and gluon contributions, which turn out to contribute to the small-$x$
eigenvector with opposite signs.\cite\bfr\

Keeping in mind that the theoretical picture is somewhat blurred, let
us turn to the data. The earlier experiment,\cite\EMCa\ which extends
down to $x=0.015$, fitted to the
{\it asymmetry} a parametrization that behaves at small $x$  as
$A_1\neath \sim {x\to 0} x^{1.12}$ which
then, assuming $F_2\neath \sim {x\to 0}{\rm const.}$, leads, by
Eq.~\aftwo, to the behavior \smallxre\ with $\alpha=0.12$. A direct fit
of the form \smallxre\ to the  last seven  data points\ref\elkara{J.~Ellis and
M.~Karliner, {\it Phys. Lett.} {\bf B213} (1988) 73} leads to
$\alpha=0.07^{+ 0.32}_{-0.42}$. Finally, a recent determination\ref\NMC{NMC
Collaboration, P.~Amaudruz et al., {\it Phys. Lett.} {\bf B295}
(1992) 159}   of $F_2$ displays a fitted behavior
$F_2\neath \sim {x\to 0}x^{-1.1}$, which, if used together with the
above parametrization of $A_1$, would lead to\cite\CR\ $\alpha=0.02$;
however if $g_1$ is determined\cite\wa\ from the measured
asymmetry\cite\EMCa\
using this form of $F_2$, the result can be still fitted with the function
\smallxre.
Quantitatively, these methods all lead to a contribution
to the first moment from the
$x<0.01$ region
\eqn\smallxfm
{\Gamma_1^{{\rm small}\, x}\equiv\int_0^{0.01}\!dx\, g_1(x)\approx 0.002,}
which is about 2\% of $\Gamma_1$ (and usually  just
included in the systematic error on the value of $\Gamma_1$).

The more
recent experimental data,\cite\SMC\ however, extend to significantly
smaller $x$ ($x_{\rm min}=0.003$), and lead to the striking result
displayed in  Fig.~3: not only there is no evidence of a fall-off
with $x$, but actually (even though with large errors) the last four
data points  display a rise at small $x$. These data have been extrapolated
 by fitting a constant value for $g_1(x)$ to the last two
points; the resulting contribution from the unmeasured small $x$
region to $\Gamma_1$ is now larger than in
Ref.~\xref\EMCb\ by a factor 2, and
adding to that the contribution from the (now measured)
$0.003\le x\le 0.01$ range
leads to a value of $\Gamma_1^{{\rm small}\, x}$ larger
than Eq.~\smallxfm\ by a factor
6 or 7. Fitting the behavior \smallxbala\
leads\cite\CRb\ to a similar result, while the very singular form
of Eq.~\smallxcloro\ gives a value which is yet larger by a factor 3,
\ie, about 20 times larger than  Eq.~\smallxfm. This alone would
increase\cite\CRb\ the value of $\Gamma_1$ by almost 25\%.

Of course such conclusions should be taken with caution,
especially since what is actually observed
experimentally is
an approximately constant behavior of the asymmetry $A_1$, and the rise is then
produced by extrapolating a parametrization to $F_2$ outside its
declared\cite\NMC\ range of validity; furthermore, perturbative evolution may
substantially
affect the small $x$ behavior\cite\bfr\ and it is unclear that a
small $x$ tail
fitted at $\langle Q^2\rangle \sim 1$~GeV$^2$ can be assumed to be
unchanged at $\langle Q^2\rangle =10.7$~GeV$^2$ where $g_1(x)$ is given.
Unfortunately the very recent SLAC data,\cite\SLAC\ which have better
statistics,  only extend down to $x_{\rm min}=0.029$; they are
consistent with the EMC/SMC data and, if fitted to a
constant,\cite\stuart\
lead to a value of $\Gamma_1^{{\rm small}\, x}$
of the same size as that of Eq.~\smallxfm.
\smallskip
\subsec{\it Determining $\Delta \Sigma$}
The singlet and nonsinglet components
of $\Gamma_1$ can be extracted  using Eq.~\firstmom.
This requires the determination of the matrix element of the triplet and octet
current, and the computation of the
coefficient functions $C(Q^2)$.

The triplet and octet
current are conserved and therefore scale independent; they can thus
be taken from any other process.
The triplet matrix element is accurately known, because
(through trivial isospin algebra)
it is equal to the axial coupling measured in nucleon $\beta$-decay:
\eqn\trip
{a_u-a_d= {G_A\over G_V}= F+D=1.2573\pm 0.0028,}
where $F$ and $D$ are octet meson decay constants in the
standard SU(3) parametrization. The octet matrix element is then
given, using SU(3) symmetry, by
\eqn\oct
{a_u+a_d-2 a_s= 3F-D=0.579\pm 0.025.}
The value in Eq.~\oct\ is obtained from the most recent fit\cite\CRb\
to octet decays; notice that these values are significantly different
(and more precise) than those used in older analyses.\cite\EMCb\

Now, $\Delta \Sigma$ is found rewriting Eq.~\firstmom\ as
\eqn\decomp
{ C_s(Q^2)\Delta\Sigma(Q^2) = 9\Gamma_1^p -{C_{NS}(Q^2)\over
2}\left(3F+D\right).}
The  two terms on the r.h.s. of Eq.~\decomp\ are roughly of the same
size (typically, $\Gamma_1\sim 0.1$), thus $\Delta \Sigma$ arises from
a large cancellation between them. In the most recent and accurate
data\refs{\SMC\SLAC} the error on $\Gamma_1$
is of order 10\%, hence it still dominates over the error on $3F+D$,
which is of order 1\%.

Knowledge of the perturbative coefficient functions in Eq.~\decomp\ has
considerably improved recently. The one-loop results have long been
known, in both the nonsinglet\cite\koda\ and
singlet\cite\kodsolo\ cases;
the nonsinglet two loop\ref\gola{S.~G.~Gorishny and S.~A.~Larin, {\it
Phys. Lett.} {\bf B172} (1986) 109} and three loop\ref\lave{S.~A.~Larin
and J.~A.~M.~Vermaseren, {\it Phys. Lett.} {\bf B259} (1991) 345}
coefficients have been determined subsequently; the singlet two loop
coefficient was recently calculated,\nref\vanne{E.~B.~Zijlstra
and W.~L.~van Neerven, {\it Nucl. Phys.} {\bf B417}
(1987) 452}\nref\lar{S.~A.~Larin, {\it Phys. Lett.}
{\bf B334} (1994) 192}
\refs{\vanne,\lar} and estimates for the next
corrections (\ie\ three loop in the singlet case, and four loop in the
nonsinglet) have  also been proposed.\ref\kata{A.~L.~Kataev, preprint
CERN-TH.7333/94 (1994)} The full results (in the $\overline{\rm MS}$
scheme) are
\eqnn\cnonsing\eqnn\csing
$$
\eqalignno{C_{NS}(Q^2)&=\Big[1-\left({\alpha_s\over\pi}\right)
-3.5833\left({\alpha_s
\over\pi}\right)^2-20.2153\left({\alpha_s\over\pi}\right)^3\cr
&\qquad\qquad-(\sim 130)\left({\alpha_s
\over\pi}\right)^4+\dots\Big]&\cnonsing\cr
C_{S}(Q^2)&=\left[1-\left({\alpha_s\over\pi}\right)
-1.0959\left({\alpha_s\over\pi}\right)^2-
(\sim 3.7)\left({\alpha_s\over\pi}\right)^3+\dots\right],&\csing\cr}
$$
where $\sim$ denotes  the estimated coefficients. The estimates
are arrived at by requiring minimal scale sensitivity (plus, in the
nonsinglet case, some guesswork), and should be
taken with care.
Two comments are in order here: first, higher loop corrections turn
out to have relatively large coefficients and are not negligible if
$Q^2\lsim 5$~GeV$^2$; also, all loop corrections go in the same
direction, namely, for given  $\Gamma_1$ the value of $\Delta \Sigma$
obtained from Eq.~\decomp\ increases as more perturbative orders are
included. The overall effect of these corrections is of order 20\%
around $Q^2\sim5$~GeV$^2$; this is large enough for the uncertainty
on $\Lambda_{\rm QCD}$ to reflect on an uncertainty on $\Delta \Sigma$
of a few per cent. This sensitivity  can be actually used to measure
$\alpha_s$ (see Sect.~6).

The scale dependence of the quantity on the l.h.s. of Eq.~\decomp\ is
still not entirely specified, because the singlet
matrix element $\Delta \Sigma$ depends on $Q^2$ (due to the
anomalous nonconservation of
the singlet axial current). Its scale dependence
starts at next to leading order; the first two nontrivial
coefficients have been
computed,\refs{\kodsolo,\lar} while the next order is estimated\cite\kata:
\eqn\dsigevol
{\Delta \Sigma (Q^2)= \left[1+{2\over
3}\left({\alpha_s\over\pi}\right)+1.2130\left({\alpha_s\over\pi}\right)^2+
(\sim 3.6)\left({\alpha_s\over\pi}\right)^3+
\dots\right]\Delta\Sigma(\infty).}
\goodbreak
\medskip
\newsec{\bf The proton experimental results and their meaning}
\nobreak
The current experimental knowledge on $\Gamma_1$ and $\Delta\Sigma$
for the proton is summarized in Table 1; the last column gives the
proton matrix element of the strange axial current,
and thus provides a measure of the Zweig rule
violation in this channel. The first two rows of the column are
obtained from the same raw data; the more recent value\cite\wa\ of
$\Delta\Sigma$
differs essentially because of the use of updated values of $F$ and
$D$ and the inclusion of higher loop corrections
(the latter, however, have
a negligible effect at this scale).
The values of
$\Delta\Sigma$  shown in the table are as given by the respective
references, except the value for the SLAC experiment,\cite\SLAC\ which
we determined using Eqs.~\decomp\ with the coefficient functions
\cnonsing-\csing\ and the decay constants \trip-\oct.
As more data become available, however, it is now more sensible to
perform a global fit of $\Delta \Sigma$ from the determinations of
$\Gamma_1$ in various experiments;\cite\guidomo\
we will discuss this in Sect.~6.
\topinsert\hfil
\vbox{\tabskip=0pt \offinterlineskip
      \def\tablerule{\noalign{\hrule}}
      \halign to 350pt{\strut#&\vrule#\tabskip=1em plus2em
                   &\hfil#\hfil&\vrule#
                   &#\hfil&\vrule#
                   &#\hfil&\vrule#
                   &#\hfil&\vrule#
                   &\hfil#&\vrule#\tabskip=0pt\cr\tablerule
      &&\omit\hidewidth Ref.~\hidewidth
      &&\omit\hidewidth $\langle Q^2\rangle $\hidewidth
      &&\omit\hidewidth $\Gamma_1^p$\hidewidth
             &&\omit\hidewidth $\Delta\Sigma$\hidewidth
             &&\omit\hidewidth $a_s$\hidewidth&\cr\tablerule
   && \xref\EMCb, \xref\SLACold && 10.7 && $0.126\pm0.018$ &&
              $0.12\pm0.17$ && $-0.19\pm0.06$&\cr\tablerule
   && \xref\wa && 10.7 && $0.126\pm0.018$ &&
              $0.14\pm0.17$ && $-0.15\pm0.06$&\cr\tablerule
   && \xref\SMC, \xref\SLACold && 10.0 && $0.142\pm0.014$ &&
              $0.27\pm 0.13$ && $-0.10\pm0.05$&\cr\tablerule
   && \xref\SLAC && 3.0 && $0.129\pm0.011$ &&
            $0.28\pm 0.11$ && \hfil$-0.10\pm0.04$\hfil &\cr\tablerule}}
\hfil\bigskip\noindent{\abstractfont
Table 1: Summary of proton
experimental results. All results hold at the given scale.}
\smallskip
\endinsert

All data points turn out to agree within errors; the
SMC result\cite\SMC\ is
substantially
larger because of the unexpectedly large values of $g_1$
in the small $x$ region, which is not covered by other experiments.
The recent, more precise (but still
preliminary) SLAC data\cite\SLAC\  also agree with previous data;
the difference in value of $\Gamma_1$ is mostly accounted for by $Q^2$
evolution.  The error on $\Delta \Sigma$ is dominated by the
uncertainty on $\Gamma_1$; this, in turn, comes in roughly equal
proportions from statistics and systematics in the EMC/SMC
experiments, whereas it is systematics--dominated in the E143
experiment,\cite\SLAC\ which has better statistics but a smaller
kinematic coverage. If the large values of $g_1$ at small $x$ of
Ref.~\xref\SMC\ are confirmed, then the small-$x$ extrapolation of
Ref.~\xref\SLAC\ will have to be corrected, since the effect on $\Gamma_1$
is then comparable to the total uncertainty.
\smallskip
\subsec{\it Theoretical interpretation}
The large violation of the Zweig rule displayed by the data of Table 1
has been variously dubbed ``spin crisis'' or ``spin puzzle''. Why this
result should be puzzling at all (after all the Zweig rule is only a
phenomenological expectation, known to fail in some channels) is
perhaps best understood from the point of view of someone who wishes
to construct a
parametrization of the polarized
quark distributions $\Delta q_i$ and gluon distribution $\Delta G$.
In leading order, $\Delta \Sigma$ is
 scale independent; also,
in the parton model
[Eq.~\naive] $\Delta \Sigma$ is just the fraction of the nucleon
helicity carried by quarks. Parton model results are modified by QCD
evolution, but since  the first moment does not evolve this
identification is retained. This is in keeping with the observation
that [Eq.~\jmel]
$\Delta\Sigma$ is the coefficient of proportionality
between the nucleon helicity and the quark axial charge, which
coincides with the quark helicity for free, massless quarks. Thus,
$\Delta \Sigma$ is  the normalization of the polarized parton
distribution
which is input to the perturbative evolution, and one would guess that
it should
just be equal to  the spin fraction
carried by quarks in quark models of the nucleon, which is typically\cite\JM\
$\sim0.6\pm 0.1$; hence the puzzle.
Of course, this is a puzzle for parton models and parametrizations,
but not for QCD, even less so for effective models of the nucleon.

The resolution of
this difficulty in the parton model\ref\AR{G.~Altarelli and G.~G.~Ross, {\it
Phys. Lett.} {\bf B212} (1988) 391}
is easy to state, and becomes apparent
when the QCD evolution equations are solved in
next to leading order.
The eigenvector of the QCD evolution equation
for the first moment turns out to be
\eqn\twid
{\Delta\Sigma=
\sum_i \int_0^1 \!dx\,\Delta\tilde q_i =\sum_i\int_0^1 \!dx\,\left(\Delta q_i
-{\alpha_s\over2\pi}\Delta g\right)}
whose  anomalous dimension is equal to that of the axial current
Eq.~\dsigevol\ (the
eigenvector remains Eq.~\twid\ even at higher orders). Now, the crucial
point is that this anomalous dimension starts at next to leading
order, which means that at leading order the combination
${\alpha_s\over2\pi}\Delta g$ is scale invariant; thus, the
gluon contribution to Eq.~\twid\ is
not asymptotically suppressed by a power of $\alpha_s$,
and the next to leading order analysis tells us that
the one loop scale invariant combination which in the parton model is
associated to $\Delta \Sigma$ is that given in Eq.~\twid,
and not the naive parton model one of Eq.~\naive. The polarized gluon
distribution whose first moment appears in Eq.~\twid\ is uniquely
defined and can be independently measured in different hard
processes,\nref\CCM{R.~D.~Carlitz, J.~C.~Collins and A.~H.~Muller,
{\it Phys. Lett.} {\bf B214} (1988) 229}\nref\AL{G.~Altarelli and
B.~Lampe, {\it Z. Phys.} {\bf C47} (1990) 315}\refs{\CCM,\AL}
even though in polarized deep-inelastic scattering it is only the
combination in Eq.~\twid\ which is measurable. It appears that
the coefficient of the
gluon contribution to Eq.~\twid\ can be changed by using different
regularizations of infrared collinear
singularities,\ref\bodqiu{G.~T.~Bodwin and J.~Qiu, {\it Phys. Rev.}
{\bf D41} (1990) 2755} thus
modifying the amount of mixing between quarks and gluons, but a
detailed analysis\ref\canto{W.~Vogelsang, {\it Z. Phys.} {\bf C50}
(1991) 275} shows that this can only be done by including soft
contributions   in a hard coefficient function.
This said,  the gluon coefficient  in
Eq.~\twid\ can still be modified by reabsorbing the gluon
contribution in a redefinition of the polarized quark distribution,
\ie\ by a change of scheme; however, the definition Eq.~\twid\ is the
unique one where $\sum_i \Delta q_i$ is conserved to all orders, as
the quark helicity must be.

The reason for these results resides in the anomalous conservation law
satisfied by the axial current
\eqn\anom
{\partial_\mu j^\mu_5=
N_f{\alpha_s\over2\pi}\tr\epsilon^{\mu\nu\rho\sigma}F_{\mu\nu}F_{\rho\sigma},}
which is exact nonperturbatively. Due to the anomaly, matrix elements
of the axial current do not measure directly the quark helicity, but
rather a combination of the quark helicity and an anomalous non-conserved
contribution;
the latter reduces perturbatively to the expression
Eq.~\twid, but can also receive a nonperturbative
contribution $\Omega$,
which need not have a partonic interpretation,
so that in general\ref\me{S.~Forte, {\it Phys. Lett.} {\bf B224} (1989)
189; {\it Nucl. Phys.} {\bf B331} (1990) 1}
\eqn\sigdefnpqcd
{\Delta \Sigma=\sum_i\int \!dx\, \Delta\tilde q_i-N_f\Omega. }
 The anomalous
contribution is due to
the helicity generated by anomalous particle
creation,\cite\me\
in analogy\ref\patel{B.~Patel, Columbia University preprint  CU-TP-599
(1993)} to electroweak baryon number
generation.\footnote*{\footnotefont\baselineskip 11 truept
The fact that the
gluon contribution can be absorbed in the quark one by a choice of
scheme, but at the expense of losing the conservation of first moment
of the polarized  quark density, corresponds to the fact that the
anomalous dimension of the axial current can be set to zero by a
finite renormalization (because it starts at two loops), but the
mixing of the axial currents with the (gluonic) anomaly operator
cannot be removed (being a one-loop effect), and the coefficient of
the mixing (the strength of the anomaly)
is fixed in a scheme independent way once the way in which $\Delta q$
renormalizes is specified. Of course, gluons will also
contribute to higher moments of $g_1$, but this contribution is
entirely scheme-dependent; the pertinent coefficient functions are
given within specific choices of scheme in
Ref.~\nref\stirge{T.~Gehrmann and W.~J.~Stirling, Durham preprint
DTP/94/38 (1994)}\xref\stirge.} In principle, the perturbative
anomalous contribution $\Delta g$ can be measured
directly\refs{\guidoer,\guidomo}, while the nonperturbative one $\Omega$
can be revealed in different processes;\ref\menp{M.~Anselmino and
S.~Forte, {\it Phys. Rev. Lett} {\bf 71} (1993) 223; {\it Phys. Lett.}
{\bf B323} (1994) 71} the former is scale dependent, as discussed
above, while the latter is not.

Regardless of whether  it is $\Delta g$ or a
nonperturbative contribution $\Omega$, a large cancellation between the
quark contribution and the anomalous gluon contribution to
$\Delta\Sigma$
 must be invoked to
explain its observed small value. This cancellation in turn can only be
explained in terms of a nonperturbative mechanism.
While effective models of the nucleon, such as the
Skyrme,\nref\BEK{S.~J.~Brodsky, J.~Ellis and
M.~Karliner, {\it Phys. Lett.} {\bf B206} (1988)
309}\nref\sche{R.~Johnson
et al., {\it Phys. Rev.} {\bf D42}
(1990) 2998\semi G.~K\"albermann,
J.~M.~Eisenberg and A.~Sch\"afer, preprint hep-ph/9409299}\refs{\BEK,\sche}
bag,\ref\rho{B.-Y.~Park et al., {\it Nucl. Phys.} {\bf A504} (1989) 829}
or Nambu--Jona-Lasinio\ref\goeke{A.~Blotz,
M.~Prasza\l owicz and K.~Goeke, Bochum preprint RUB-TPII-41/93} models can
easily be made to accommodate the observed values of $\Delta \Sigma$,
$F$ and $D$ (but at the expense of  intoducing extra
free parameters\footnote\dag{\footnotefont\baselineskip 11 truept
For
instance, the pure Skyrme model in the $N_f\to\infty$
limit predicts\cite\BEK\ $\Delta \Sigma=0$ and cannot
fit $F$ and $D$ accurately, but can be made to fit all the data by
introducing corrections in ${1\over N_f}$ such as those due to
coupling to vector mesons.\cite\sche}), in these models it
is impossible to separate  quark and gluon contributions to
observed quantities.
On the other hand, a cancellation between $\Delta q$ and $\Omega$ in
Eq.~\sigdefnpqcd\ can be explicitly shown to take place due to instanton-like
configurations in the QCD vacuum,\ref\meshu{S.~Forte and
E.~V.~Shuryak, {\it Nucl. Phys.} {\bf B357} (1991) 153}
however it is hard to go
beyond model calculations.

A rather different understanding of the small value of $\Delta
\Sigma$ is arrived at in $t$-channel approaches, where the nucleon
matrix element of the axial current is computed coupling the current
to composite operators that correspond to physical bound states,
which then couple irreducibly to the nucleon.
The small value of $\Gamma_1$ is then
due to a strong nonperturbatively induced scale
dependence,\ref\bal{R.~D.~Ball. {\it Phys. Lett.} {\bf B266}
(1991) 473} or to the smallness of the relevant bound state propagator,
which can be expressed using an exact Ward identity in terms of the
topological susceptibility of the vacuum,\ref\SV{G.~Shore and
G.~Veneziano, {\it Nucl. Phys.} {\bf
B381} (1992) 3} and estimated using QCD sum
rules,\ref\NSV{S.~Narison,
G.~Shore and G.~Veneziano, preprint CERN-TH.7223/94
(1994)} with results in excellent agreement with
experiment.
These approaches, which rely only on the
nonperturbative QCD dynamics, imply that the smallness of $\Gamma_1$
depends on the structure of the  axial current, and is unrelated to
the specific target which is being considered;\cite\NSV\ their relation to the
parton approach, however, is not immediate.
\smallskip
\subsec{\it Polarized parton parametrizations}
The problem of constructing a parametrization of polarized parton
distributions is actually not an academic one. Earlier proposals, based on
the idea of relating polarized distributions to unpolarized ones
through a dilution factor\ref\CarlK{R.~Carlitz and J.~Kaur, {\it
Phys. Rev. Lett.} {\bf D38} (1977) 673} lead necessarily to results in
disagreement\cite\EMCa\ with the data; however, once the anomalous
gluon contribution Eq.~\twid\ is taken into account, the data can be
fitted phenomenologically without giving up the idea that polarized
and unpolarized quark distributions should have the same gross features,
either within the same approach,\ref\CKfoll{D.~de~Florian et
al.,
{\it Phys. Lett.} {\bf B319}, (1993) 285; preprint hep-ph/9408363} or
by constructing a parametrization from scratch.\ref\otherpar{K.~Sridhar and
E.~Leader,
{\it Phys. Lett}, {\bf B295} (1992)283\semi J.~Bartelski and S.~Tatur,
preprint hep-ph/9409204} Of course the data can also be described
without anomalous contribution but assuming a large amount of Zweig rule
violation (which is absent in the unpolarized case).

As more data
become available, however, it is now possible to construct polarized
parton distributions which incorporate all the constraints from
perturbative QCD [Eqs.~\courule, \smallxbro]
as well as the general constraint from Regge theory that the small $x$
behavior be isosinglet.\cite\stirge\
Imposing the constraints at a starting scale
of $Q^2=4$~GeV$^2$ the data can then be fitted rather accurately,
consistently including the scale dependence at leading
order,
and imposing cuts to keep higher twist corrections
under control.
The fit can be performed assuming the polarized sea quark
distributions to be entirely generated by perturbative
evolution; $\Delta G$ turns out to be largely unconstrained,
and can be chosen to have different large $x$ behaviors. If the
gluon is assumed to behave as a constant at small $x$, then the fit
to the proton data\refs{\SLACold,\EMCb,\SMC}
favors a singular quark of the form Eq.~\smallxre\ (which dominates
$g_1$ at small $x$), with $\alpha\approx-0.55\pm0.15$. The neutron
structure function can then be predicted, the data not being good
enough to constrain it seriously. Eventually, it should be possible to
test these parametrization directly: while $\Delta G$ can be
measured typically by photon-gluon fusion,\cite\guidoer\
$\Delta q$ can in principle
be measured for sea  and valence flavors separately, by tagging
final state hadrons; preliminary results\ref\wis{SMC Collaboration,
W.~Wi\'slicki, preprint hep-ex/9405012} are consistent with the
results of the fit to the proton data.\cite\stirge\
\goodbreak
\medskip
\newsec{\bf Polarized scattering on neutrons}
\nobreak
A measurement of $g_1^n$ provides the determination of an independent
linear combination of polarized quark distributions.
Thus, neutron experiments, besides giving independent theoretical
information on $\Delta \Sigma$, also offer the possibility of
measuring directly the nonsinglet polarized structure function
$g_1^p-g_1^n$,
whose first moment is just [from Eq.~\firstmom]
\eqn\gamnonsing
{\Gamma_1^p-\Gamma_1^n={1\over 6} C_{NS}(Q^2) (a_u-a_d).}
A comparison of
the data with the known value of $a_u-a_d$ [Eq.~\trip] and the
computed $C_{NS}$ [Eq.~\cnonsing] thus  allows a direct test of
 isospin in this channel, as well as of the predicted
scale dependence. However, neutron experiments are somewhat subtler,
both because nuclear effects have now to be taken into
account, and because they generally have
a more restricted kinematic coverage than  proton experiments.

The neutron structure function can only be measured by scattering on
nuclear targets. Specifically, deuterium\refs{\SMCn,\SLAC} and
${}^3$He targets\ref\SLACn{E142 Collaboration, P.~L.~Anthony et al., {\it
Phys. Rev. Lett.} {\bf 71} (1993) 959}  have been used. In the
former case, assuming additivity, one obtains a determination of
$g_1^p+g_1^n$; in the latter case, the two protons are mostly in an $S$-state
so that the spin of the  nucleus is carried by the neutron.
In both cases, however, there are complications due to  nuclear
structure. Taking into account that the deuteron can be in $D$ wave
with probability $\omega_D$ one has\cite\SMCn\
\eqn\gamman
{\Gamma_1^p+\Gamma_1^n={2\Gamma_1^p\over 1-1.5\omega_D}.}
The ${}^3$He
wave function has in general 10 components; a simplified description
in terms of a three-component wave function leads
to\ref\ciofi{C.~Ciofi
degli Atti, E.~Pace and G. Salm\'e, Istituto Superiore di Sanit\`a (Rome)
preprint INFN-ISS-93-3}
\eqn\gammahe
{\Gamma^n_1=(1.15\pm0.02)\Gamma_1^{He}+(0.057\pm0.009)\Gamma_1^p,}
which however only  holds  in the Bjorken limit, and could be
significantly corrected at finite $Q^2$; the error involved is
still negligible compared to present day experimental accuracy.
Further nuclear effects are due to Fermi motion (which for the
deuteron is
estimated\ref\tok{M.~V.~Tokarev, {\it Phys. Lett.} {\bf B318} (1994)
559} to lead to a sizable correction to $g_1$, up to perhaps 10\%),
and to shadowing and
antishadowing.\ref\shad{N.~N.~Nikolaev and V.~I.~Zakharov, {\it Phys. Lett.}
{\bf B55} (1975) 397\semi S.~J.~Brodsky and H.~J.~Lu, {\it Phys. Rev.
Lett.} {\bf 64} (1990) 1342} No systematic investigation of the latter
effects is available yet.

The problems related to $Q^2$ evolution and $x$ extrapolation discussed
in  Sect.~3.1 are even more serious in neutron experiments.
An example is the uncertainty involved in the large
$x$  behavior
 of the deuterium data\cite\SMCn: if this is
 determined assuming a valence quark model\cite\CR\ at large $x$, rather than
fitting the data directly, the value\cite\SMCn\ of $\Gamma_1^n$
increases  by a
factor two;\cite\CR\
however, the data  are sufficiently uncertain that the two
results are still compatible within errors.
The problems related to small $x$
are shown in Fig.~4, which displays the extrapolation of the E142 data,
performed assuming a
Regge-like behavior $A_1^n\neath\sim{x\to0} x^{1.2}$,\cite\SLACn\
compared with
an extrapolation of neutron data obtained by combining the SMC
data\cite\SMCn\ with the EMC proton data,\cite\EMCb\ and then fitting a
power behavior Eq.~\smallxre\ to the last data point. These two
extrapolations lead to values of $\Gamma_1^n$ which differ by over
30\%.

The issue of scale dependence is particularly serious here because all
neutron data are taken at very low $Q^2$. This not
only means that the (calculable) effects of QCD evolution due to
difference in scale between $x$ bins are now rather sizable (the
correction on the first moment is of order 10\%),\cite\anr\ but also,
that higher twist effects can be important. If these effects
are included, then, for instance, the scale dependence of the
nonsinglet first moment
in Eq.~\gamnonsing\ is modified as
\eqn\gamht
{\Gamma_1^p-\Gamma_1^n={1\over 6} \left[C_{NS}(a_u-a_d)+{c_{HT}\over
Q^2}\right].
}
The value of $c_{HT}$ in Eq.~\gamht\ has been the subject of
considerable controversy: QCD sum rules lead\ref\bb{I.~I.~Balitsky,
V.~M.~Braun and A.~V.~Koleschnichenko, {\it Phys. Lett.} {\bf B242}
(1990) 245; Erratum {\bf B318} (1993) 648} to $c_{HT}=-0.09\pm0.06$,
but a different estimate based on the same method\ref\RR{G.~G.~Ross
and R.~G.~Roberts, {\it Phys. Lett.} {\bf B322} (1994) 425} has
$c_{HT}=-0.15\pm0.02$; a bag model computation\ref\jiun{X.~Ji and
P.~Unrau, MIT preprint CTP-2232 (1993)} even leads to a result with the
opposite sign, $c_{HT}=0.16$. The sum rule method  appears to be
self-consistent in that a different sum
rule\ref\newht{E.~Stein et al., Frankfurt preprint UFTP 366/1994
(1994)}
leads essentially to the same result as Ref.~\xref\bb; the theoretical
uncertainty is perhaps of order 50\%.\cite\newht\ The
scale dependence of the isotriplet $\Gamma_1$ is displayed in Fig.~5;
higher twist effects are included according to Eq.~\gamht\ with
$c_{HT}=-0.1$. This shows that the magnitude of higher twist
correction is comparable to that of three loop corrections, but
smaller than the uncertainty on $\Lambda$.

The neutron experiments are summarized in Table 2. The result
of Ref.~\xref\wa\ (shown in Fig.~4) is obtained by putting together
the neutron data of
Ref.~\xref\SLACn\
with the values of $g_1^n$ obtained combining $g_1^p+g_1^n$ determined
from the deuterium experiment\cite\SLACn\ with proton
data.\refs{\EMCb,\SLACold}
The values of $\Delta \Sigma$ and $a_s$ are as quoted by the
respective references, except the entry corresponding to the
preliminary
result of Ref.~\xref\SLAC, which we determined using the weak decay
constants Eqs.~\trip-\oct,
and
the coefficient functions \cnonsing-\csing.
\topinsert\hfil
\vbox{\tabskip=0pt \offinterlineskip
      \def\tablerule{\noalign{\hrule}}
      \halign to 350pt{\strut#&\vrule#\tabskip=1em plus2em
                   &\hfil#\hfil&\vrule#
                   &#\hfil&\vrule#
                   &#\hfil&\vrule#
                   &#\hfil&\vrule#
                   &\hfil#&\vrule#\tabskip=0pt\cr\tablerule
      &&\omit\hidewidth Ref.~\hidewidth
      &&\omit\hidewidth $\langle Q^2\rangle $\hidewidth
      &&\omit\hidewidth $\Gamma_1$\hidewidth
             &&\omit\hidewidth $\Delta\Sigma$\hidewidth
             &&\omit\hidewidth $a_s$\hidewidth&\cr\tablerule
   && \xref\SMCn && 4.7 && $\Gamma_1^d=0.023\pm0.025$ &&
              $0.06\pm0.25$ && $-0.21\pm0.08$&\cr\tablerule
   && \xref\SLACn && 2.0 && $\Gamma_1^n=-0.022\pm 0.011$ &&
              $0.57\pm0.11$ && $-0.01\pm0.06$&\cr\tablerule
   && \xref\wa  && 5.0 && $\Gamma_1^n=-0.055\pm0.025$ &&
              $0.24\pm 0.23$ && $-0.11\pm0.08$&\cr\tablerule
   && \xref\SLAC && 3.0 && $\Gamma_1^d=0.044\pm0.005$ &&
            $0.35\pm 0.05$ && \hfil$-0.08\pm0.03$\hfil &\cr\tablerule}}
\hfil\bigskip\noindent{\abstractfont
Table 2: Summary of neutron and deuteron experimental results. All results
hold at the given  scale;
$\Gamma_1^d\equiv(\Gamma_1^p+\Gamma_1^n)/2$}
\smallskip
\endinsert

Again, all experiments agree within errors. The error
on the SMC
data\cite\SMCn\ is mostly statistical, whereas the E142-E143
data,\refs{\SLACn,\SLAC} which have generally better statistics (but
smaller kinematic coverage), have approximately equal
statistical and systematic uncertainty. The
significantly larger values of $\Gamma^n_1$ found by the SLAC
experiments\refs{\SLACn,\SLAC}  appear to be
mostly due to the small-$x$ region,
as shown by the fact  that the values of $\Gamma_1^n$ of
Refs.~\xref\SLACn\
and \xref\wa\ differ  essentially because of
 the contribution from this region (notice however that they
are compatible within errors).
If the trend displayed by the SMC
data at small $x$ (Fig.~4) is confirmed, the small $x$ extrapolation of
the more precise data\refs{\SLAC} would have to be corrected
accordingly.
\goodbreak
\medskip
\newsec{\bf Summary: the data and sum rules}
\nobreak
The determinations of $\Gamma_1$
with proton, neutron, and deuterium
targets can now be compared with each other and with
QCD expectations; the results, evolved to a common scale using the perturbative
coefficient functions of Sect.~3.2 to highest available order (omitting
higher
order estimates  and higher twist corrections), are displayed in Fig.~6.

First, the isotriplet first moment can be checked
against the prediction of Eq.~\gamnonsing\ (Bjorken sum
rule\ref\bj{J.~D.~Bjorken, {\it Phys. Rev.} {\bf 148} (1966) 1467}),
thereby checking isospin in this channel. As  is apparent from
Fig.~6, the prediction is perfectly verified within current
experimental errors, which means that it is verified to 10\% accuracy.
If, as the experimental precision improves, a violation of the
sum rule
within these bounds were found, it would not necessarily
signal a
breakdown of perturbative QCD: on the one hand, it has been
argued\ref\halp{I.~Halperin, Tel Aviv preprint TAUP-2127-93 (1993)}
that isospin breaking in meson-nucleon couplings could lead
to a violation of the Bjorken
sum rule up to 20\%; also, the anomalous contribution $\Omega$
[Eq.~\sigdefnpqcd] to the singlet matrix element $\Delta\Sigma$,
due to its infrared sensitivity,
could develop\ref\sfbj{S.~Forte, {\it Phys. Lett.} {\bf B309} (1993)
174}  a nonsinglet component (triggered by SU(2)
violation in light quark masses) of up to 10\%.

Then, a common value of $\Delta \Sigma$ can be extracted. This is best
done\cite\guidomo\ by fitting the expression \firstmom\ to the data
with fixed
values of the weak decay constants, and $\Delta\Sigma(Q^2=\infty)$
left as a free parameter. The result, obtained from all the data of
Tables 1 and 2 (excluding the entries corresponding to Ref.~\xref\wa)
is\cite\guidomo\
\eqn\dsigwa
{\Delta\Sigma(Q^2=\infty)=0.31\pm0.04.}
This corresponds to  $a_s(Q^2=\infty)=-0.097\pm0.018$: the
parton-based expectation that $a_s$ should be compatible with zero
(Ellis-Jaffe sum rule\ref\elja{J.~Ellis and R.~L.~Jaffe, {\it Phys.
Rev.}
{\bf D9} (1974) 1444}) is thus violated by several standard deviations,
with the theoretical implications discussed in Sect.~4.1.

Finally, the considerable scale sensitivity of the first moments
(compare Fig.~5), especially in the $Q^2\lsim5$~GeV$^2$ region,
can be used to measure $\alpha_s$.\nref\elkaras{J.~Ellis and
M.~Karliner, preprint CERN-TH-7324/94 (1994)}\refs{\elkaras,\guidomo}
The simplest way of
doing this is to take advantage of the fact that both the
scale dependence and the normalization of the isotriplet
combination in Eq.~\gamnonsing\ are accurately predicted, so that
comparing
an  experimental determination of $\Gamma_1^p-\Gamma_1^n$
at finite $Q^2$ to the asymptotic value Eq.~\trip\
gives immediately a determination  of
$\alpha_s$. Including all known
perturbative corrections and estimates of higher loops (but not higher
twist corrections), and using $\Gamma_1^n$
 of Ref.~\xref\SLACn\ (as reanalysed in Ref.~\xref\EK) and
$\Gamma_1^p$ of Ref.~\xref\SLAC\ gives\cite\elkaras\
\eqn\als
{\alpha_s(M^2_Z)=0.122^{+0.005}_{-0.009},}
in good agreement with the current world average, and with
surprisingly small error. This error, however, does not
include the uncertainty related to the higher twist correction:
if it is taken into account
using  Eq.~\gamht\ with
$c_{HT}=-0.12
\pm0.06$ (according to the sum rule
estimate discussed in Sect.~5), then\cite\elkaras\
$\alpha_s(M^2_Z)=0.118^{+0.007}_{-0.014}$.
A
determination which includes all the available information can be arrived
at fitting  Eq.~\firstmom\ to all the available data,
with  both $\Delta\Sigma$ and $\alpha_s$ left as free parameters;
the result is\cite\guidomo\ (neglecting again higher twist effects)
\eqn\totfit
{\Delta\Sigma(Q^2=\infty)=0.33\pm0.04;\quad\alpha_s(M^2_Z)=0.125\pm0.006.}
This result is once more surprisingly accurate, and in good agreement with
Eq.~\als. It would be interesting to also fit
simultaneously $\alpha_s$, $\Delta \Sigma$, and the octet combination
$a_u+a_d-2a_s$, but current data are not accurate
enough.\refs{\elkaras,\guidomo}

In conclusion, combining all available results leads to a consistent
picture, in excellent agreement with the predictions of perturbative
QCD, and with interesting implications for the nucleon structure.
While the improving quality of experimental data requires a
more sophisticated theoretical analysis, it is now possible to test
our understanding of several subtle perturbative QCD effects, and to
start coping with the nonperturbative effects which determine the
structure of the nucleon.
\nobreak
\medskip
\noindent{\bf Acknowledgements:} I wish to thank
 B.~L.~Ward for his warm hospitality in
Tennessee;  G.~Altarelli and L.~Stuart for discussions,
R.~Ball for a critical reading of the manuscript, F.~Close and
M.~Einhorn for comments,
and especially
G.~Ridolfi for considerable help in analysing the data.
\goodbreak
\medskip

\immediate\closeout\rfile\writestoppt
\bigskip
\noindent{{\bf References}}\bigskip{\frenchspacing%
\parindent=20pt\escapechar=` \input refs.tmp\vfill\eject}\nonfrenchspacing
\vfill
\eject
\centerline{\bf Figure Captions}
\bigskip
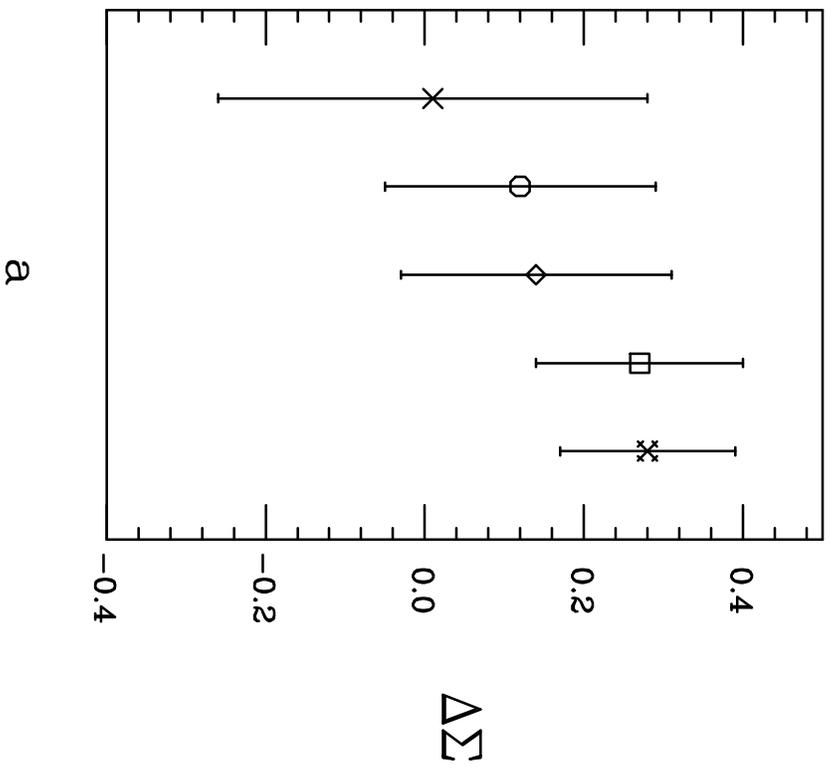
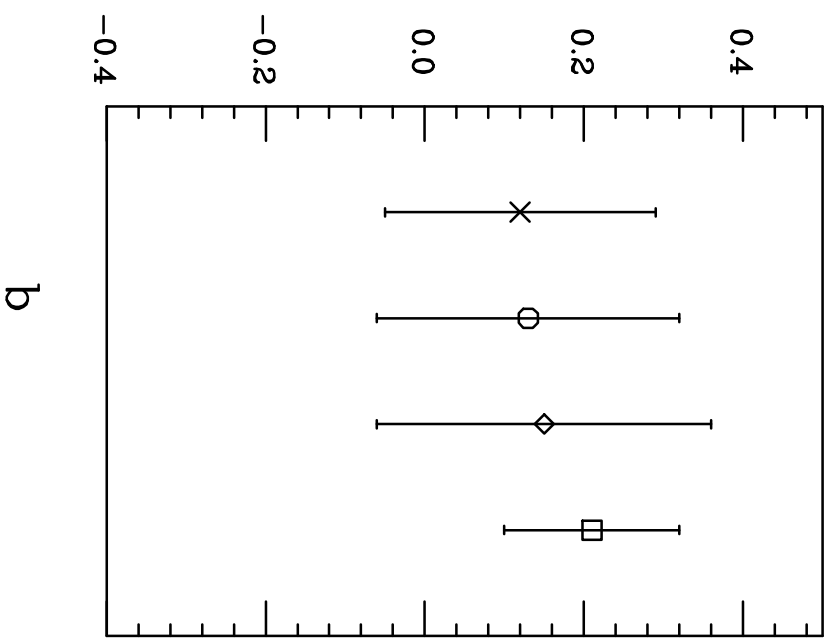
\item{\bf Fig.~1:} Experimental determinations of
$\Delta \Sigma^p$ (at the scale of the respective experiments, see
Tab.1 below):
a) cross, EMC (1988);\cite\EMCa circle, EMC
(1989);\cite\EMCb
diamond, world average
(1994);\cite\wa
square, SMC (1994);\cite\SMC
star, E142 (1994).\refs{\SLAC,\stuart}
b) Theoretical
reanalysis of the 1989  experiment\cite\EMCb: cross, published
value\cite\EMCb; circle, Jaffe and Manohar (1990);\cite\JM
diamond, Ellis and Karliner (1993);\cite\EK square, Close and
Roberts (1993).\cite\CR
\medskip
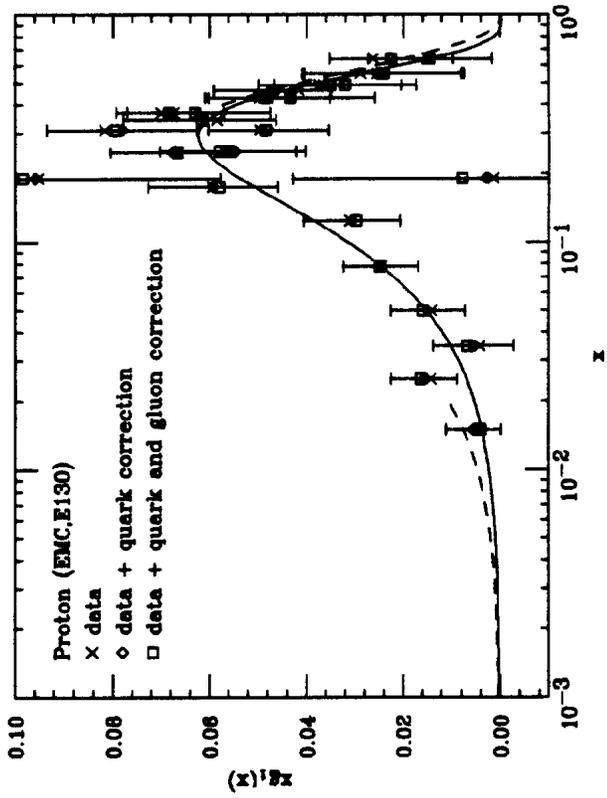
\item{\bf Fig.~2:} (from Ref.~\xref\anr): a) Scale dependence of the
asymmetry $A_1(x, Q^2)$, compared to the data. The crosses indicate data from
Ref.~\xref\SLACold\ and the squares data from
Ref.~\xref\EMCb; the low  (high) curves correspond to the lower (upper)
edge of the $x$-bin, the solid (dashed) lines corresponds to a maximal
polarized gluon (no polarized gluon). b) Effect on the
data\refs{\SLACold, \EMCb}
of the scale dependence correction.
\medskip
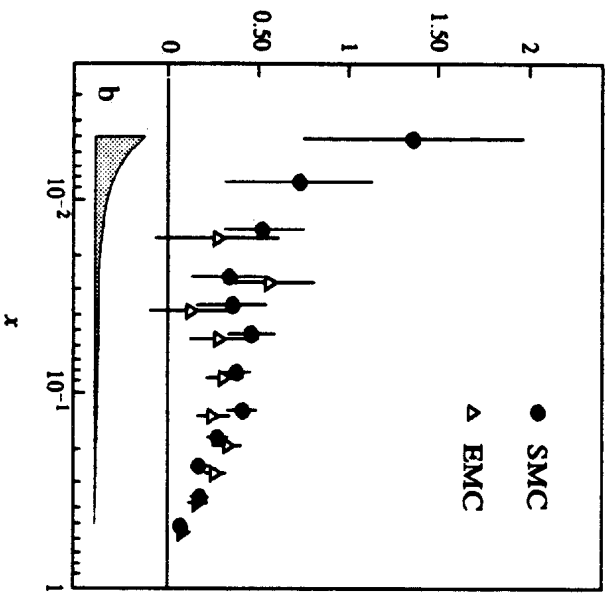
\item{\bf Fig.~3:} a)(from Ref.~\xref\EMCb) Experimental data for
$g_1^p(x)$ at $\langle Q^2\rangle =10.7$~GeV$^2$ (filled dots,
EMC data;\cite\EMCb\
open circles, SLAC data\cite\SLACold); the dashed line is the small--$x$
extrapolation. b)(from Ref.~\xref\SMC) The data of Fig3a (here as
triangles) compared to the new data.\cite\SMC
\medskip
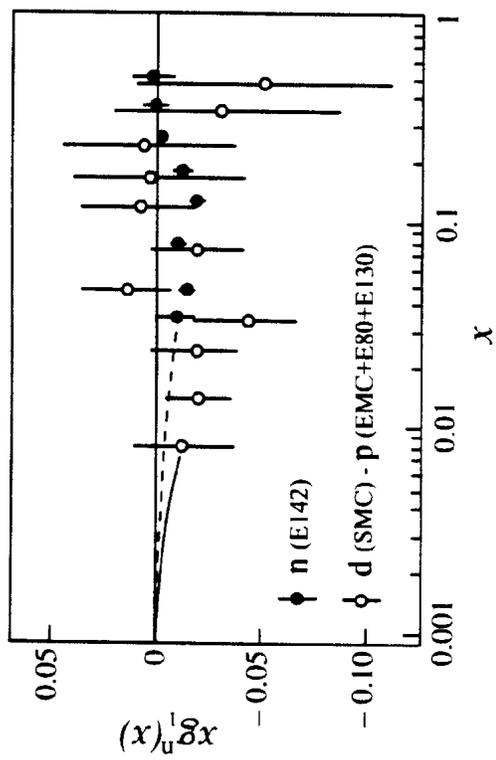
\item{\bf Fig.~4:}(from Ref.~\xref\wa) The E142\cite\SLACn\ and
SMC/EMC\refs{\SMC,\EMCb} data with the respective small $x$
extrapolations (dashed: E142; solid: SMC/EMC)
\medskip
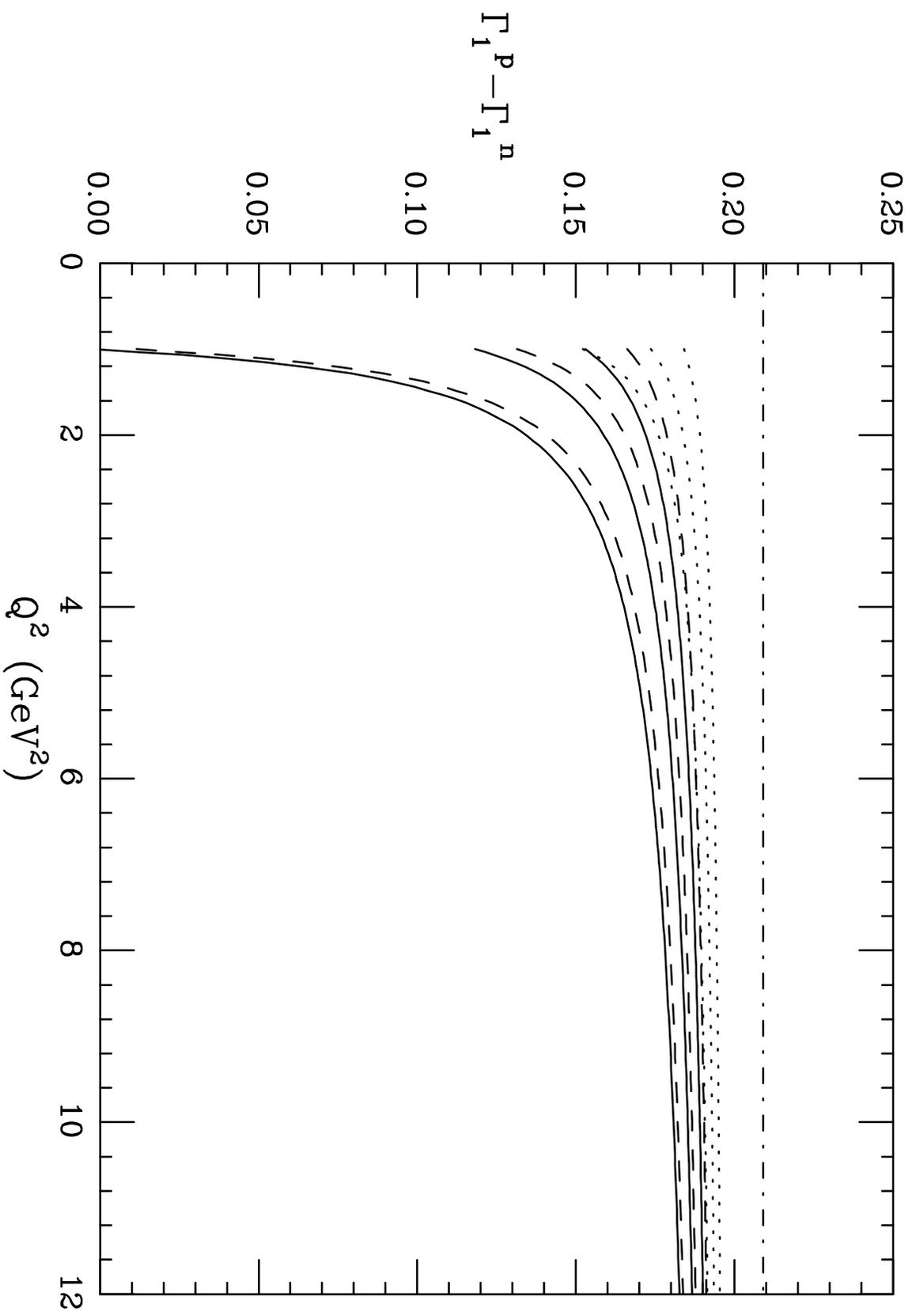
\item{\bf Fig.~5:} (adapted from Ref.~\xref\anr) Scale
dependence of the isotriplet first moment Eq.~\gamnonsing. Dashdot:
tree-level result ($C_{NS}=1$);
dotted: Leading order; dashed: three loops; solid: with higher twist.
The three sets of curves correspond to the values $\Lambda=267,\;383,\;
509$~MeV.
\medskip
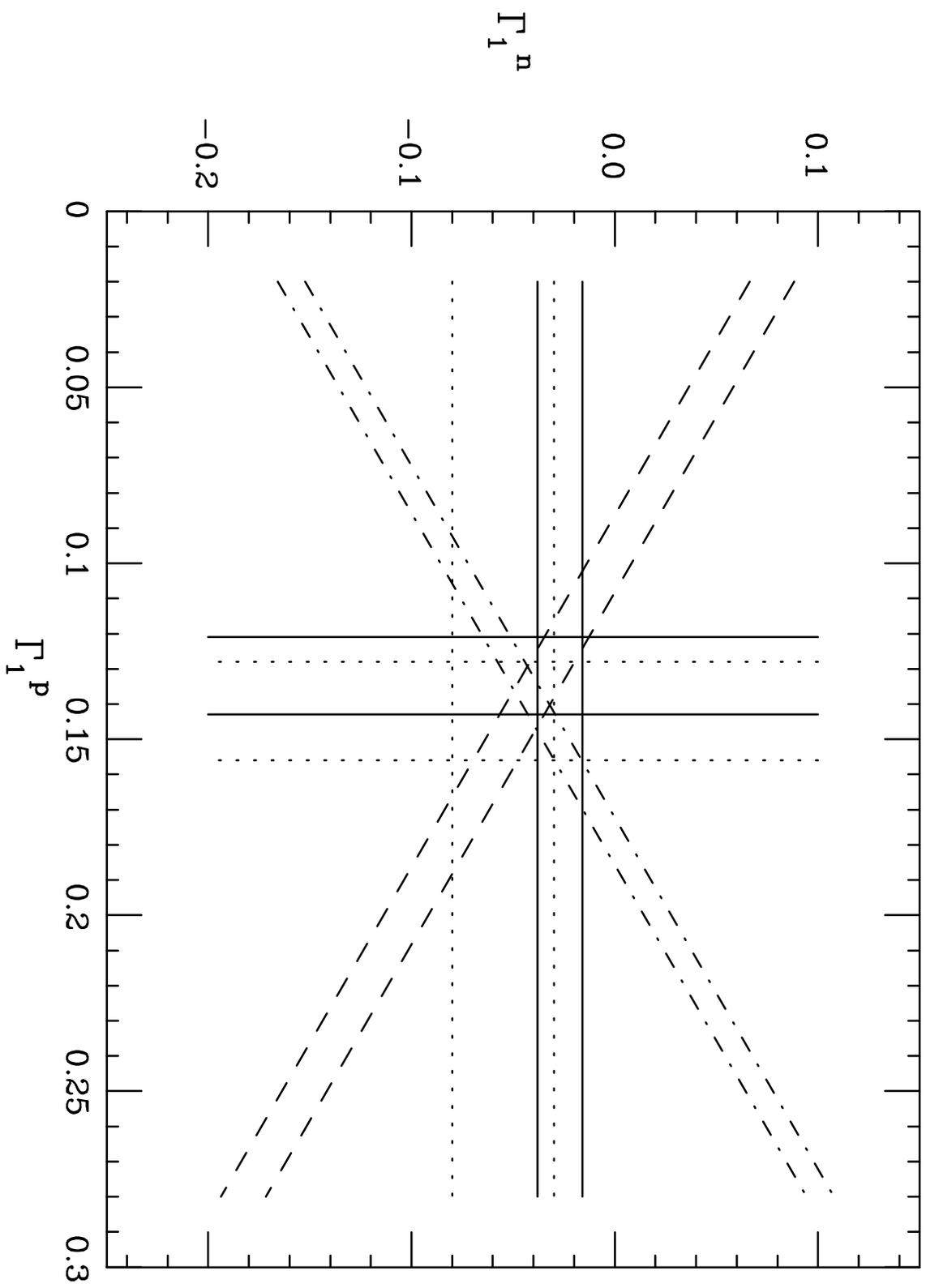
\item{\bf Fig.~6:} Comparison of experimental determinations of
$\Gamma_1$ for proton,\cite\SLAC neutron\cite\SLACn\ and (dashed)
deuteron.\cite\SLAC\ The dotted lines indicate alternative
proton\cite\EMCb\ and neutron\cite\wa\ determinations. The Bjorken sum
rule prediction Eq.~\gamnonsing\ is also shown (dotdashed).
All results have been
evolved to $Q^2=5$~GeV$^2$.
\medskip
\vfill
\eject
\bye